\documentclass
[superscriptaddress,secnumarabic,nobibnotes,aps,prd,showkeys,noshowpacs,twocolumn]{revtex4-2}%
\usepackage{graphics}
\usepackage{graphicx}
\usepackage{color}
\usepackage{epsf}
\usepackage{bm}
\usepackage{amsmath,amssymb,amsfonts,mathrsfs,amsthm}
\usepackage{latexsym}
\usepackage{enumerate}
\usepackage{comment}
\usepackage{hyperref}
\usepackage{epstopdf}%

\def\x{\mbox{\rm x}}
\def\y{\mbox{\rm y}}

\begin{document}
\title{Hidden Symmetries in Deformed Very Special Relativity}
\author{N. Dimakis}
\email{nsdimakis@scu.edu.cn, nsdimakis@gmail.com}
\affiliation{Center for Theoretical Physics, College of Physics, Sichuan University, Chengdu 610064, China}

\begin{abstract}
We study particle dynamics in a space-time invariant under the $DISIM_b(2)$ group - the deformation of the $ISIM(2)$ symmetry group of very special relativity. We find that the Lorentz violation leads to the creation of higher order (hidden) symmetries, which are connected to those broken at the space-time level. Through the perspective of the conserved quantities of the special relativistic case, the Lorentz violation is linked to specific noncommutative relations in phase-space.
\end{abstract}
\maketitle

\section{Introduction}

The subject of Lorentz violation has a long history in theoretical physics and is motivated by different arguments in a broad spectrum of theories (string theory, loop quantum gravity, noncommutative geometry etc. \cite{paper1,paper2,paper3,paper4}). In this work, we are interested in the symmetry structure of the particle dynamics in the deformed version of very special relativity (VSR), which incorporates such a violation. In this context, we reveal the existence of higher order symmetries, which are connected to breaking Lorentz invariance.

VSR was introduced by Cohen and Glashow in \cite{Cohen}. The basic idea is that Lorentz symmetry is not a fundamental symmetry of nature but rather this role is reserved for one of its proper subgroups. One such realization consists of taking the four-parameter similitude $SIM(2)$ subgroup of the Lorentz group, together with the translations, in order to form the eight-dimensional $ISIM(2)$ group. In \cite{GiGoPo}, it was demonstrated that the $ISIM(2)$ group admits a physically acceptable deformation which was called $DISIM_b(2)$, with $b$ being a nonzero dimensionless parameter whose value needs to be very small ($b<10^{-26}$). This deformed algebra is compatible with a line element belonging to a Finsler type of geometry, which was initially introduced by Bogoslovksy \cite{Bogo1,Bogo2}. Relativistic particle dynamics in Finsler geometry is often used to model dispersion relations emanating from Lorentz violation in field theory \cite{mech1,mech2,mech3,mech4}.

In our case, the geometry is characterized by the Bogoslovsky-Finsler line element
\begin{equation} \label{Bogolineel}
  ds^2 = g_{\mu\nu}dx^\mu dx^\nu \left[\frac{\left(\ell_\mu dx^\mu\right)^{2}}{-g_{\mu\nu}dx^\mu dx^\nu}\right]^b ,
\end{equation}
with $g_{\mu\nu}$ being the flat space-time metric and $\ell$ a covariantly constant, null, future directed Killing vector of $g_{\mu\nu}$. The $ds^2$ of \eqref{Bogolineel} is a homogeneous function of degree $2$ in the $dx^\mu$, which places it in the class of Finsler geometries. Of course, when the parameter $b$ is zero, the typical pseudo-Riemannian line element of special relativity is recovered. The $b\rightarrow0$ limit however is quite subtle at the level of the finite symmetry transformations since the preferred direction introduced with $\ell$ in \eqref{Bogolineel} persists in those relations (see \cite{Dhasmana}).

In what follows, we work in light cone coordinates, $x^\mu=(v,u,\x,\y)$, in which the pseudo-Riemannian metric $g$ involved in \eqref{Bogolineel}, is
\begin{equation} \label{Minkuv}
  g =g_{\mu\nu} dx^\mu dx^\nu = 2 du dv + \delta_{ij} dx^i dx^j,
\end{equation}
while the null vector becomes $\ell =\ell^\mu \partial_\mu= \partial_{v}$. Of course, the context of the theory can be generalized by adopting curved space-time metrics (for studies in pp-wave geometries, see \cite{Fuster,HorvGib}).

Throughout this work, greek indices cover the whole spectrum of space-time variables, the $i,j$ are reserved for the $\x-\y$ plane, while the $u,v$ subscripts are used to denote the components in the corresponding null direction. Greek indices are raised and lowered with the four-dimensional metric \eqref{Minkuv}, while for $i,j$, the $\delta_{ij}$ is used for this purpose. Finally, in our conventions, $ds^2<0$ and $g<0$ for timelike distances.

Unlike the pseudo-Riemannian line element \eqref{Minkuv}, which is invariant under the Poincar\'e $\mathfrak{iso}(3,1)$ algebra, the Bogoslovsky-Finsler line element is invariant under the transformations generated by
\begin{subequations} \label{eightsym}
\begin{equation} \label{sevensym}
  T_\mu = \partial_{\mu}, \quad B_{ij} = x_i \partial_{j} - x_j  \partial_{i} , \quad B_{v i} = u \partial_{i} - x_i \partial_{v}
\end{equation}
and
\begin{equation} \label{Nbsym}
  N = (1+b) v \partial_{v} + (b-1) u \partial_u +  b x^i \partial_i .
\end{equation}
\end{subequations}
These vectors are the generators of the $DISIM_b(2)$ group and elements of the corresponding $\mathfrak{disim}_b(2)$ algebra, whose nontrivial Lie brackets are:
\begin{equation}
  \begin{split}
   & [N, T_v]= -(1+b)T_v, \quad [N, T_u] = (1-b) T_u, \\
   & [N,T_i]=-b T_i, \quad [N, B_{v i}] = -B_{v i} \\
   & [T_u,B_{vi}]= T_i, \quad [T_i,B_{12}] = \epsilon_{ij}T_j, \\
   & [T_i,B_{vj}]= - \delta_{ij} p_v, \quad [B_{vi},B_{12}]=\epsilon_{ij}B_{vj} ,
  \end{split}
\end{equation}
where $\epsilon_{ij}$ is the antisymmetric tensor with $\epsilon_{12}=1$.

The vectors \eqref{sevensym} are of course also part of $\mathfrak{iso}(3,1)$, spanned by $T_\mu$ and $B_{\mu\nu}=x_\mu \partial_{\nu} - x_\nu  \partial_{\mu}$, which are the isometries of the pseudo-Riemannian metric $g$. The three remaining vectors of the Poincar\'e algebra which leave invariant $g$ but fail to be symmetries of $ds^2$ are represented by:
\begin{equation} \label{missing}
  B_{u v} = v \partial_{v} - u \partial_u , \quad B_{ui} = v \partial_i -x_i \partial_u  .
\end{equation}
The vector $N$ is the one carrying the deformation parameter $b$. When the latter is zero, the $N|_{b=0}$ is identified with the $B_{u v}$ vector which together with the seven vectors from \eqref{sevensym} form the generators of the $ISIM(2)$ group.

We have thus the following setting: i) the eight-dimensional $\mathfrak{disim}_b(2)$ algebra spanned by \eqref{eightsym}, which leaves invariant the $ds^2$ and ii) the ten-dimensional Poincar\'e algebra of \eqref{sevensym} and \eqref{missing}, which are isometries of $g$. We shall refer to the latter as the $b=0$ case throughout the manuscript due to $g=ds^2|_{b=0}$.

Apart from this fundamental difference at the symmetry level though, there exists a striking resemblance between line elements \eqref{Bogolineel} and \eqref{Minkuv}. According to a theorem proven by Roxburgh \cite{Rox}, certain classes of Finslerian spaces have the property of reproducing the exact same geodesics as Riemannian metrics. The Bogoslovsky-Finsler line element \eqref{Bogolineel} belongs to this class, and even though it has a distinct symmetry structure compared to the $g$ of \eqref{Minkuv}, it has its extrema exactly on the same trajectories as the latter. This very interesting property is the motive of this work. We seek to find the deeper connection between the two systems and what happens to the symmetries and the corresponding integrals of motion that are broken in the Bogoslovsky-Finsler case by the Lorentz symmetry violation.

\section{The hidden symmetries}

In order to dig deeper into the symmetry structure of the problem we study the motion of a particle of mass $m$ in the Bogoslovsky-Finsler spacetime characterized by line element \eqref{Bogolineel}. The corresponding action can be written as $S =-m\int\! \sqrt{-ds^2}=\int\!\tilde{L}d\tau$ (we work in units $c=1$), where $\tilde{L} = - m \left(\ell_\mu \dot{x}^\mu\right)^b \left(-\dot{x}^\mu \dot{x}_\mu \right)^{\frac{1-b}{2}}$. The dot denotes differentiation with respect to the parameter along the curve which we symbolize with $\tau$. It is more convenient however to use instead of  $\tilde{L}$ the equivalent Lagrangian
\begin{equation}\label{Lag2}
  L = -\frac{1}{2 e} \dot{u}^{2b} \left(-\dot{x}^\mu \dot{x}_\mu \right)^{1-b} -e \frac{m^2}{2},
\end{equation}
where $e=e(\tau)$ is an auxiliary degree of freedom called the einbein \cite{Brink} and in which $\ell_\mu dx^\mu=du$ was used. Lagrangian $L$ corresponds to a well-defined Hamiltonian which can be obtained through the Dirac-Bergmann algorithm \cite{Dirac,AndBer}; in contrast to $\tilde{L}$ which is a function homogeneous of degree 1 in the velocities and thus its Hamiltonian is bound to be identically zero \cite{Lanczos}.

The equivalence of the two Lagrangians is straightforward upon calculation of the equations of motion. The Euler-Lagrange equation for the degree of freedom $e$, i.e. $\frac{\partial L}{\partial e}=0$, leads to
\begin{equation} \label{con}
  e^2 = \frac{1}{m^2} \dot{u}^{2 b} \left(-\dot{x}^\mu \dot{x}_\mu \right)^{1-b},
\end{equation}
which is the constraint relation of the system. Substitution of $e$ from \eqref{con} into $L$ maps the latter to $\pm \tilde{L}$, depending on the sign in front of the square root that one takes when solving \eqref{con}. The use of \eqref{con} inside the second order equations of motion, reduces them to
\begin{equation} \label{ELred}
  \ddot{u} = \dot{u} \frac{\ddot{v}}{\dot{v}}, \quad \ddot{x}^i = \dot{x}^i \frac{\ddot{v}}{\dot{v}}
\end{equation}
with $v(\tau)$ remaining an arbitrary function through which the parametrization invariance of the system is expressed. This last set \eqref{ELred} is equivalent to the one obtained from the Euler-Lagrange equations of $\tilde{L}$, which  - as a constrained system of equations - can be solved algebraically with respect to just three accelerations; the solution being \eqref{ELred}. More important than the equivalence at the level of the equations is that $\tilde{L}$ and $L$ admit the same symmetries. The transformations induced by \eqref{eightsym} leave invariant both Lagrangians. The situation is similar to what happens in the Riemannian case, where $b=0$, and $L|_{b=0}$ becomes the quadratic equivalent of the square root Lagrangian $\tilde{L}|_{b=0}=-m\sqrt{-\dot{x}^\mu\dot{x}_\mu}$; either one can be used to study a geodesic problem.

The ``time'' gauge choice $v(\tau)=\tau$ leads to $e=$constant, which corresponds to the typical affine parametrization.  As expected by the theorem proven by Roxburgh \cite{Rox}, eqs. \eqref{ELred} are the same as those for the motion in Minkowski space-time where $b=0$. The difference of the two systems rests in the association of the constants of integration with the physical parameters $m$ and the now nonzero $b$ through the constraint equation \eqref{con}.

As we already mentioned, the symmetry structure of the Bogoslovsky-Finsler line element and consequently of Lagrangians $\tilde{L}$ and $L$ is quite different from that of the $b=0$ case. However, the fact that the same set of second order equations provides solutions in both cases signifies that there are conservation laws to be accounted for in the $b\neq0$ case since it admits a smaller symmetry group.

Our study on conserved charges can be better expressed in the Hamiltonian formulation. To this end the Dirac-Bergmann algorithm for constrained systems \cite{Dirac,AndBer} is applied. We refrain from presenting details on the theory of constrained systems and we refer to relevant textbooks \cite{Dirac2,Sund}. The resulting Hamiltonian constraint reads
\begin{equation}\label{Ham}
  H = -\frac{(1-b)^{b-1}}{(1+b)^{1+b}} p_v^{-2 b} \left(-p_\mu p^\mu \right)^{1+b} + m^2 = 0,
\end{equation}
where $p_\mu = \frac{\partial L}{\partial \dot{x}^\mu}$. Equation \eqref{Ham} leads to the dispersion relation: $p_\mu p^\mu =2 p_u p_v + p_i p^i = -m^2 (1-b^2) \left(\frac{p_v^2}{m^2(1-b)^2} \right)^{\frac{b}{1+b}}$, which was first presented in \cite{GiGoPo}. The equality to zero in \eqref{Ham} holds on mass shell and in the formalism of constrained systems it is referred to as a weak equality \cite{Dirac}.

As is expected by Noether's theorem, the symmetries \eqref{eightsym} of the Bogoslovsky-Finsler line element generate linear in the momenta integrals of motion which are
\begin{equation} \label{integrals}
  \begin{split}
    & I_\mu=p_\mu, \quad I_{ij} = x_i p_{j} -x_j p_{i}, \quad  I_{vi} = u p_i - x_i p_v, \\
    & I_N = (1+b) v p_v + (b-1) u p_u + b x_i p_{i},
  \end{split}
\end{equation}
and of course have the property of commuting with $H$. Apart from the above conserved charges, we may notice that the following quantities, which are rational functions in the momenta, are also conserved:
\begin{subequations}\label{extraI}
\begin{align}
  I_{uv} & =  v p_v - u p_u + \frac{b}{1+b} u \frac{p_\mu p^\mu}{p_v}   \\
  I_{ui} & = v p_i - x_i p_u  + \frac{b}{1+b} x_i \frac{p_\mu p^\mu}{p_v} .
\end{align}
\end{subequations}
It is straightforward to check that truly $\{I_{uv},H\}=0=\{I_{ui},H\}$, where $\{,\}$ are the usual Poisson brackets. An interesting point about this new charges is that they look like $b$-distorted ``boosts'', since for $b=0$ they fall to the linear Minkowski space charges generated by vectors \eqref{missing}.

The fact that the quantities $I_{uv}$, $I_{ui}$ possess a nonlinear dependence on the momenta means that they are not generated by space-time vectors like \eqref{missing}, but of what is called higher order or hidden symmetries of the Lagrangian. The components of such symmetry generators depend also on derivatives of the coordinates. For more information on these types of symmetries we refer to \cite{Olver,Cariglia} (maybe the most famous hidden symmetry in physics is the one associated with the Carter constant for the geodesic motion in Kerr space-time \cite{Carter}). The symmetry generators of  $I_{uv}$ and $I_{ui}$ are
\begin{subequations} \label{symvecs}
\begin{align}
  X_{uv} & = - u \partial_u + \left(v+ \frac{b}{1-b} \frac{\dot{x}^\mu \dot{x}_\mu}{\dot{u}^2} u  \right) \partial_v \\
  X_{ui} & = - x_i \partial_u + \frac{b}{1-b}\frac{\dot{x}^\mu \dot{x}_\mu}{\dot{u}^2}  x_i  \partial_v + v \partial_i \; .
\end{align}
\end{subequations}
It is easy to check that if we take into account the relation,
\begin{equation} \label{momtovelratio}
  \frac{p_\mu p^\mu}{p_v^2}= \frac{(1+b)\dot{x}^\mu \dot{x}_\mu}{(1-b)\dot{u}^2},
\end{equation}
then we directly obtain the $I_{uv}$ and $I_{ui}$ of \eqref{extraI} through calculating the inner product of the vectors with the momentum, i.e. $I_{uv}=(X_{uv})^\mu p_\mu$, $I_{ui}=(X_{ui})^\mu p_\mu$.

The transformations which the $X_{uv}$ and $X_{ui}$ induce in the space of $(x^\mu, \dot{x}^\nu)$, leave invariant both Lagrangians $L$ and $\tilde{L}$. The ensuing transformations are recovered by extending the symmetry vectors $X=X^\mu\partial_\mu$ in the space of the first derivatives, i.e. $X^{[1]}= X + \dot{X}^\mu \frac{\partial}{\partial \dot{x}^\mu}$. The $X^{[1]}$ are called the first prolongations of the vectors $X$ \cite{Olver}. These extended vectors have a wide application in the study of symmetries in dynamical systems, many prominent examples being in cosmology \cite{Cap,Andr}. The components $\dot{X}^\mu$ of course contain now accelerations, but these can be substituted through the use of the second order Euler-Lagrange equations \eqref{ELred} and thus, obtain a transformation rule confined in the space $(x^\mu, \dot{x}^\nu)$. The invariance of the Lagrangians can be simply shown by noting that $X_{uv}^{[1]}(L)=0=X_{ui}^{[1]}(L)$ modulo eqs.  \eqref{ELred}.

The $X_{uv}$ and $X_{ui}$ of \eqref{symvecs} are not space-time vectors. Nevertheless, they can be linked to such upon the solution space. First, we note that the ratio, $R=\frac{\dot{x}^\mu \dot{x}_\mu}{\dot{u}^2}$, appearing in \eqref{symvecs} is itself a constant of motion: $\dot{R}=0$,
by virtue of the Euler-Lagrange equations \eqref{ELred}. If we use the constraint equation \eqref{con}, together with the first integral of motion from eq. \eqref{integrals}, which reads in the velocity phase space
\begin{equation}
  I_v = p_v = \frac{\partial L}{\partial \dot{v}} = \frac{1}{e}(1-b) \frac{\dot{u}^{2 b+1}}{\left(-\dot{x}^\mu \dot{x}_\mu\right)^{b}}  =\pi_v,
\end{equation}
then we may write the on mass shell value of $R$ in terms of the parameters $m$ and $\pi_v$ (we use $\pi_v$ as the on mass shell constant value of the momentum $p_v$),
\begin{equation}\label{ratio}
  R = \frac{\dot{x}^\mu \dot{x}_\mu}{\dot{u}^2} = -\left[\frac{(1-b)^2 m^2}{\pi_v^2}\right]^{\frac{1}{1+b}}.
\end{equation}
In view of the last expression, we produce a reduced (on mass shell) version of vectors \eqref{symvecs} as
\begin{subequations} \label{redvecs}
\begin{align} \label{redvec1}
  \xi_{uv} & = - u \partial_u + \left[v - b (1-b)^{\frac{1-b}{1+b}} \left(\frac{m^2}{\pi_v^2}\right)^{\frac{1}{1+b}} u \right] \partial_v \\
  \xi_{ui} & = - x_i \partial_u - b (1-b)^{\frac{1-b}{1+b}} \left(\frac{m^2}{\pi_v^2}\right)^{\frac{1}{1+b}}  x_i  \partial_v + v \partial_i \; .
\end{align}
\end{subequations}
These $\xi_{uv}$, $\xi_{ui}$ are indeed space-time vectors and when we set $b=0$ in \eqref{redvecs}, we recover the missing Minkowski space symmetries $B_{uv}$ and $B_{ui}$ appearing in \eqref{missing}. We can even use them to write simplified, linear in the momenta, on mass shell equivalent expressions for the \eqref{extraI}. Those are $Q_{uv}=(\xi_{uv})^\mu p_\mu$ and $Q_{ui}=(\xi_{ui})^\mu p_\mu$, which commute with the Hamiltonian when the relations $H=0$ and $p_v=\pi_v$ are enforced.

The $\xi_{uv}$, $\xi_{ui}$ close an algebra together with the seven vectors \eqref{sevensym}. It is however a trivial deformation of the Poincar\'e algebra, i.e. it is the same algebra expressed in different coordinates. This can be directly seen by noticing that the set of these ten vectors are isometries of the following \underline{flat} space metric
\begin{equation} \label{gtild}
  \bar{g} = g- b  (1-b)^{\frac{1-b}{1+b}} \left(\frac{m^2}{\pi_v^2} \right)^{\frac{1}{1+b}} \ell_\mu \ell_\nu dx^\mu dx^\nu ,
\end{equation}
where $g$ is given by \eqref{Minkuv}. The space-time transformation for which $\bar{g}\mapsto g$ serves as an algebra automorphism and makes the corresponding vectors (up to linear combinations) assume the usual expressions leading to the typical Poincar\'e algebra. Equation \eqref{gtild} implies that the metric $\bar{g}$ is disformally related to $g$. Disformal transformations were initially defined by Bekenstein as generalizations to conformal transformations \cite{Bekenstein} (for applications see also \cite{Lobo1,Lobo2}).

We need to stress here that the $\xi_{uv}$, $\xi_{ui}$ of \eqref{redvecs} are not themselves symmetry vectors but the reduced expressions of the higher order symmetries given by the $X_{uv}$, $X_{ui}$ in \eqref{symvecs}. The symmetry group of the system (when referring to space-time vectors) is the $DISIM_b(2)$; it is interesting to see however that the effect of the parameter $b$, and the violation of Lorentz invariance does not result in a complete elimination of the broken symmetries from the system. They are converted into the higher order symmetries, $X_{uv}$, $X_{ui}$, which reduce on mass shell to expressions given by the original Killing vectors distorted appropriately by $b$, the $\xi_{uv}$ and the $\xi_{ui}$ respectively. Similar distortions of broken symmetries have emerged in a different setting involving proper conformal Killing vectors in the case of the motion of a massive particle in Riemannian pp-wave space-times \cite{ppwaves}: for example, it is well known that in the case of null geodesics the proper conformal Killing vectors (CKVs) generate integrals of motion. However this property is lost when one considers a massive particle (in a sense the presence of the mass $m$ breaks these symmetries). In \cite{ppwaves} it was shown that the proper CKVs still contribute by producing conservation laws under a similar mass dependent distortion. Here, in a Finsler geometry, we see it happening at the level of Killing vectors whose symmetry property is broken by the introduction of the nonzero parameter $b$.

We can actually extend the hidden symmetries we encountered here by also using proper CKVs together with the necessary distortions. However this goes outside the scope of this letter where we mainly want to focus to the effect of the distortion on the Lorentz symmetry vectors. Just as an example though, we mention that by taking the following linear combination of the symmetry $N$ given in \eqref{Nbsym} together with the vector $\xi_{uv}$ of \eqref{redvec1}
\begin{equation}
  \begin{split}
 \xi_h & = \frac{1}{b} N - \frac{1-b}{b} \xi_{uv} \\
 & = \left[2 v + \left(\frac{(1-b)^{2} m^{2}}{\pi^{2}}\right)^{\frac{1}{1+b}} u \right] \partial_v + x^i \partial_i ,
  \end{split}
 \end{equation}
we obtain the distortion of the homothecy, and the $\xi_h |_{(b=0=m)}$ is the homothetic vector of $g$. We can now use $\xi_h$ to write the (reduced) linear conserved quantity $Q_h = (\xi_h)^\mu p_\mu$ and even go backwards with the substitution of \eqref{ratio} in conjunction with \eqref{momtovelratio} to finally express the original integral of motion which is generated by a higher order Noether symmetry and which reads
\begin{equation}
 I_h = 2 v p_v + x^i p_i - \frac{1-b}{1+b} \frac{p_\mu p^\mu}{p_v} u  .
\end{equation}
A direct calculation shows that $\{I_h,H\}=0$.

\section{Noncanonical coordinates}

As we mentioned, the mapping $\bar{g}(u,v,x^i)\mapsto g(U,V,x^i)$ allows us to obtain the Poincar\'e algebra generators out of linear combinations of the ten vectors of \eqref{sevensym} and \eqref{redvecs}. In order to avoid any confusion we use the $(u,v,x^i)$ for the original coordinates and $(U,V,x^i)$ for those after the transformation. The mapping $(u,v,x^i) \mapsto (U,V,x^i)$ we need is
\begin{equation} \label{utoU}
  v = V + \frac{b\left(1-b\right)^{\frac{1-b}{1+b}}}{2}  \left(\frac{m^2}{\pi_v^2}\right)^{\frac{1}{1+b}} U,
\end{equation}
while the rest of the coordinates remain unchanged, i.e. $u=U, x^i=x^i$. Under the aforementioned coordinate change we have
\begin{equation} \label{lcomb}
  \begin{split}
    & T_u + \frac{b\left(1-b\right)^{\frac{1-b}{1+b}}}{2}  \left(\frac{m^2}{\pi_v^2}\right)^{\frac{1}{1+b}} T_v  \mapsto T_U:= \partial_U , \\
    & \xi_{ui}  - \frac{b\left(1-b\right)^{\frac{1-b}{1+b}} }{2} \left(\frac{m^2}{\pi_v^2}\right)^{\frac{1}{1+b}} B_{vi}  \mapsto B_{U i} :=V \partial_i -x_i \partial_U,
  \end{split}
\end{equation}
while for the rest we get: $\xi_{uv}\mapsto B_{UV}$, $B_{vi}\mapsto B_{V i}$, $B_{ij}\mapsto B_{ij}$, $T_v\mapsto T_V$ and $T_i\mapsto T_i$. Thus, the Poincar\'e algebra corresponding to the symmetries of the flat metric $g(U,V,x^i)= 2 dU dV + dx^i dx_i$ is completely recovered.


What would be of interest though, is to acquire a mapping that connects the full symmetry generated charges \eqref{extraI} of the Bogoslovsky-Finsler line element to the conserved quantities emerging from the Poincar\'e algebra of the flat space metric $g(U,V,x^i)$. For this we use as a guide transformation \eqref{utoU} and substitute in it the constants $m$, $\pi_v$ with respect to their phase space dynamical equivalents from \eqref{ratio} and \eqref{momtovelratio}. After this process, we obtain the following noncanonical transformation from $U,V$ to $u,v$ variables
\begin{equation} \label{noncanonical}
  \begin{split}
    U=u,  \quad V = v + \frac{b }{2 (1+b) } \frac{p_\mu p^\mu}{p_v^2} u, \\
    p_U = p_u -\frac{b}{2 (1+b)} \frac{p_\mu p^\mu}{p_v}, \quad p_V =p_v,
  \end{split}
\end{equation}
with the $x^i, p_i$ remaining unchanged. It can be seen that this transformation maps the linear conserved charges generated by the Poincar\'e symmetry algebra of $g(U,V,x^i)$ to the higher order ones of the Bogoslovsky-Finsler metric. To make it specific, with the use of \eqref{noncanonical}, we obtain the correspondence,
\begin{align}
 I_U & := p_U = I_u - \frac{b}{2(1+b)} \frac{p_\mu p^\mu}{p_v^2} I_v  \\
 I_{Ui}& := V p_i - x_i p_U  = I_{ui} + \frac{b}{2(1+b)} \frac{p_\mu p^\mu}{p_v^2} I_{vi} .
\end{align}
The rest are mapped directly to the corresponding quantities, i.e. $I_V:=p_V=I_v$, $I_{UV}:=V p_V-U p_U=I_{uv}$, $I_{Vi}=I_{vi}$ while the $I_i$, $I_{ij}$ are obviously not affected by the transformation \eqref{noncanonical}.

Due to \eqref{noncanonical} being a noncanonical transformation, the relevant $I$ in the $(U,V)$ variables do not in general commute with the resulting Hamiltonian obtained by using \eqref{noncanonical} in \eqref{Ham}. This can be achieved by introducing the fundamental bracket relations which are implied by transformation \eqref{noncanonical}. We thus write a new bracket $[\;,\;]_b$ whose nonzero values in the $(U,V)$ coordinates are
\begin{equation} \label{noncomm}
  \begin{split}
  & [U,V]_b = \frac{b\, U}{(1+b) p_V}, \quad  [U,p_U]_b = \frac{1}{1+b},  \\
  & [V,x_i]_b = -\frac{b\, U p_i}{(1+b)p_V^2},  \quad [V,p_U]_b = \frac{b\left(p_U p_V+ p_i p^i\right)}{(1+b)p_V^2},  \\
  & [V,P_V]_b=1,  \quad  [x_i,p_U]_b = -\frac{b p_i}{(1+b)p_V},  \quad [x^i,p_j]_b=\delta^i_j .
  \end{split}
\end{equation}
These are calculated with the help of the Poisson brackets in the $(u,v)$ coordinates, e.g. $[U,V]_b=\{U(u,v,p),P_U(u,v,p)\}$. It is easy to verify that the Jacobi identity is satisfied by \eqref{noncomm}. We notice the space-time non-commutativity introduced since $V$ does not commute with either $U$ or $x^i$.

With relations \eqref{noncomm}, the transformed Hamiltonian (eq. \eqref{Ham} under use of \eqref{noncanonical}) commutes with the Poincar\'e charges in the $U,V$ variables, and it reproduces the correct equations of motion in these coordinates. In addition, the structure of the $\mathfrak{disim}_b(2)$ algebra is not affected. For example, if we take the transformed $I_N$ conserved charge which reads in the $U,V$ coordinates,
\begin{equation}
  I_N = (1+b) \left(V p_V - U p_U \right) + b x^i p_i - b U \frac{p_i p^i}{p_V},
\end{equation}
we immediately calculate $[p_U, I_N]_b = (1-b) p_U$ and $[p_V, I_N]_b = -(1+b) p_V$; exactly as we have $\{p_u,I_N\} = (1-b) p_u$ and $\{p_v,I_N\} = -(1+b) p_v$ with the $I_N$ of \eqref{integrals} in the original $u,v$ variables.

We thus see that from the point of view of the symmetries of special relativity, the effect of the Lorentz violation with the introduction of a Finslerian line element can be simulated by the introduction of noncommutative coordinates in space-time and in particular ones that satisfy relations \eqref{noncomm}. The parameter $b$ in this context signifies the deviation from the Poisson bracket formalism.

The compatibility of noncommutative space-times with the original notion of VSR and the symmetry groups that it involves has been explored previously in \cite{Jabbari} for noncommutative matrices $\theta^{\mu\nu}$ that depend solely on space-time coordinates, i.e., $[x^\mu,x^\nu]= \theta^{\mu\nu}(x)$. In our case the correspondence we make involves noncommutativity in all of phase space. Noncommutative expressions including momenta through a different approach have been used before \cite{Das} in order to reproduce the $\mathfrak{disim}_b(2)$ algebra from $B_{\mu\nu}$ and $p_\mu$. However, the algebra given there, introduces a nonzero bracket between the null momenta, $\{p_u,p_v\}\neq 0$, which is not in agreement with the original $\mathfrak{disim}_b(2)$ algebra.

\section{Conclusion}

We demonstrated that the broken symmetries due to Lorentz violation in the scenario of the deformed VSR take part in the creation of higher order symmetries from which integrals of motion that are rational functions in the momenta emerge. With a noncanonical transformation we mapped this enhanced set of conserved quantities to the usual integrals of motion of the free particle in special relativity.

We used this correspondence to define new brackets that make the charges of the Minkowskian motion commute with the Hamiltonian of the Bogoslovsky-Finsler case in the new variables. The brackets give rise to noncommutative relations in space-time. It is interesting to note that in \eqref{noncomm} the bracket, $[U,p_U]_b = \frac{1}{1+b}$, yields a different result from $[V,p_U]_b=[\x,p_{\x}]_b=[\y,p_{\y}]_b=1$. If indeed the origin of $b$ is quantum mechanical, as mentioned in \cite{GiGoPo}, the quantum analogues of the above expressions imply a possible anisotropy in the measurement of $\hbar$; the relative difference being of the order $\frac{\Delta \hbar}{\hbar} \sim b$. Given that $b<10^{-26}$, this is well below the current capacity of measuring the Planck constant, which yields relative uncertainties in a range from $10^{-9}$ to $10^{-6}$ \cite{Planck1,Planck2}. For a review on the subject of temporal and spatial variations in fundamental constants, we refer to \cite{Uzan}.

This work is concentrated on free particle dynamics; a very interesting approach for future research would be to study if the effect that is present here, i.e., the broken symmetries changing character and leading to a higher order structure, is some generic property which can be encountered in more complicated Lagrangians and especially in field theory.

\begin{acknowledgments}
This work is supported by the Fundamental Research Funds for the Central Universities, Sichuan University Full-time Postdoctoral Research and Development Fund No. 2021SCU12117
\end{acknowledgments}

\end{document}